\documentclass[12pt]{article}
 \usepackage{epsfig}
 \def\be{\begin{equation}}
 \def\ee{\end{equation}}
 \def\bea{\begin{eqnarray}}
 \def\eea{\end{eqnarray}}
 \usepackage{graphicx}

 \catcode`\@=11
 \def\lsim{\mathrel{\mathpalette\@versim<}}
 \def\gsim{\mathrel{\mathpalette\@versim>}}
 \def\@versim#1#2{\vcenter{\offinterlineskip
 \ialign{$\m@th#1\hfil##\hfil$\crcr#2\crcr\sim\crcr } }}
 \catcode`\@=12

 \catcode`@=12
\begin{document}
 \thispagestyle{empty}
 \begin{flushright}
 UCRHEP-T603\\
 Nov 2020\
 \end{flushright}
 \vspace{0.6in}
 \begin{center}
 {\LARGE \bf Dark Matter from $SU(6) \to SU(5) \times U(1)_N$\\}
 \vspace{1.2in}
 {\bf Ernest Ma\\}
 \vspace{0.2in}
{\sl Department of Physics and Astronomy,\\ 
University of California, Riverside, California 92521, USA\\}
\end{center}
 \vspace{1.2in}

\begin{abstract}\
Matter and dark matter are unified under the framework of 
$SU(6) \to SU(5) \times U(1)_N$.  A dark-matter candidate is possible, 
not because it is stable, but because it has a very long lifetime, in 
analogy to that of the proton in theories of grand unification.  A specific 
example is presented.
\end{abstract}

 \newpage
 \baselineskip 24pt

\noindent \underline{\it Introduction}~:~
The existence of dark matter appears to be indisputable~\cite{bh18}.  
Instead of taking it for granted as an {\it ad hoc} addition to the Standard 
Model (SM) of quarks and leptons, a more fundamental question may be asked 
as to its relationship with visible matter.  One possible 
answer is that both belong to the same organizing symmetry, such as $SO(10)$, 
but are distinguished by a marker symmetry such as $U(1)_\chi$~\cite{m18,m19} 
in $SO(10) \to SU(5) \times U(1)_\chi$.  Fermions and scalars which are odd 
and even under $U(1)_\chi$ belong to the visible sector, whereas fermions 
and scalars which are even and odd under $U(1)_\chi$ belong to the dark 
sector.  They are distinguished by $(-1)^{Q_\chi+2j}$ where $j$ is the 
particle's spin.  The lightest dark particle is assumed to be neutral 
and is stable because of this odd-even symmetry. 

Another possible answer is that there is no marker symmetry and both 
visible and dark matter coexist in multiplets of an organizing symmetry 
such as $SU(6)$~\cite{b12,m13}, but the dark-matter candidate itself has a 
very long lifetime, just as the proton has a very long lifetime in 
theories of grand unification.  A specific complete model of $SU(6) \to SU(5) 
\times U(1)_N$ is presented here for the first time, where a dark fermion 
decays to SM particles through a superheavy gauge boson.

\noindent \underline{\it SU(6) Unification of Visible and Dark Matter}~:~
Consider the extension of the well-known $SU(5)$ model~\cite{gg74} of grand 
unification to $SU(6)$ with $SU(6) \to SU(5) \times U(1)_N$.
Although one family of fundamental fermions under $SU(5)$ is contained 
in the anomaly-free combination of $5^*$ and $10$, the analogous case 
for $SU(6)$~\cite{ikn77,y78} is two $6^* = (5^*,-1) + (1,5)$ and one 
$15 = (10,2) + (5,-4)$. Let
\begin{equation}
6^*_{F1}  = \pmatrix {d^c \cr d^c \cr d^c \cr e \cr \nu \cr N_1}, ~~~ 
6^*_{F2}  = \pmatrix {D^c \cr D^c \cr D^c \cr E^- \cr E^0 \cr N_2}, ~~~ 
15_F = \pmatrix {0 & u^c & u^c & -u & -d & -D \cr -u^c & 0 & u^c & -u & -d 
& -D \cr u^c & -u^c & 0 & -u & -d & -D \cr u & u & u & 0 & -e^c & -E^+ \cr 
d & d & d & e^c & 0 & -\bar{E}^0 \cr D & D & D & E^+ & \bar{E}^0 & 0}.
\end{equation}
Their $SU(3)_C \times SU(2)_L \times U(1)_Y \times U(1)_N$ assignments are 
listed in Table 1.  Note that all are left-handed.
\begin{table}[tbh]
\centering
\begin{tabular}{|c|c|c|c|c|c|}
\hline
fermion & $SU(3)_C$ & $SU(2)_L$ & $U(1)_Y$ & $U(1)_N$ \\
\hline
$d^c$ & $3^*$ & 1 & 1/3 & $-1$ \\ 
$(\nu,e)$ & 1 & 2 & $-1/2$ & $-1$ \\ 
$N_1$ & 1 & 1 & 0 & $5$ \\ 
\hline
$D^c$ & $3^*$ & 1 & 1/3 & $-1$ \\ 
$(E^0,E^-)$ & 1 & 2 & $-1/2$ & $-1$ \\ 
$N_2$ & 1 & 1 & 0 & $5$ \\ 
\hline
$(u,d)$ & 3 & 2 & 1/6 & 2 \\ 
$u^c$ & $3^*$ & 1 & $-2/3$ & 2 \\
$e^c$ & 1 & 1 & 1 & 2 \\ 
\hline
$D$ & 3 & 1 & $-1/3$ & $-4$ \\ 
$(E^+,\bar{E}^0)$ & 1 & 2 & 1/2 & $-4$ \\ 
\hline
\end{tabular}
\caption{Fermion content of $SU(6) \to SU(5) \times U(1)_N$ model.}
\end{table}

The scalar sector consists of 
\begin{itemize}
\item (1) $84_S = (5,1) + (45,1) + (24,-5) + (10,7)$ which breaks the 
$U(1)_N$ of $SU(6)$ along $(1,1,0,-5)$ of $(24,-5)$ [$v_1$] to 
$SU(3)_C \times SU(2)_L \times U(1)_Y$, 
\item (2) $21^*_S = (15^*,-2) + (5^*,4) + (1,10)$ which breaks the $U(1)_N$ 
of $SU(6)$ along $(1,10)$ [$v_2$] to $SU(5)$,
\item (3) $35_S = (1,10) + (5,6) + (5^*,-6) + (24,0)$ which breaks $SU(6)$ 
along $(1,1,0,0)$ [$v_3$] from $(24,0)$ to 
$SU(3)_C \times SU(2)_L \times U(1)_Y \times U(1)_N$,  
\item (4) $15^*_S$ which breaks $SU(2)_L \times U(1)_Y$ along $(1,2,-1/2,4)$ 
[$v_4$] from $(5^*,4)$ to $U(1)_Q$,
\item (5) $84'_S = (5,1) + (45,1) + (24,-5) + (10,7)$ which breaks 
$SU(2)_L \times U(1)_Y$ along $(1,2,1/2,1)$ 
[$v_5$] from $(45,1)$ to $U(1)_Q$.
\end{itemize}
The $SU(6)$ symmetry is first broken to 
$SU(3)_C \times SU(2)_L \times U(1)_Y \times U(1)_N$ by $v_3$ at a high 
scale.  Subsequent breaking of $U(1)_N$ is by 
$v_{1,2}$ and $SU(2)_L \times U(1)_Y$ by $v_{4,5}$. 
In addition, the $Z_2$ discrete symmetry is imposed so that $6^*_{F1}$ and 
$84'_S$ are odd and the other multiplets are even.  This symmetry is obeyed 
by all dimension-four terms in the Lagrangian, but is softly broken by the 
trilinear scalar coupling $84_S \times 15^*_S \times 84'_S$, which contains 
the $v_1 v_4 v_5$ term.  

\noindent \underline{\it Fermion Masses}~:~
Because of the $Z_2$ symmetry, $6^*_{F1}$ is distinguished from $6^*_{F2}$.  
Hence only $6^*_{F2} \times 15_F$ transforms as $84_S$ and only 
$6^*_{F1} \times 15_F$ transforms as $84'_S$.  Thus $D^c D$ and 
$E^- E^+ + E^0 \bar{E}^0$ masses are proportional to $v_1$, whereas 
$d^c d$ and $e e^c$ masses are proportional to $v_5$.  Similarly,  
both $6^*_{F1} \times 6^*_{F1}$ and $6^*_{F2} \times 6^*_{F2}$ transform as  
$21_S^*$ and $15_S^*$, so that $N_{1,2}$ have Majorana masses proportional 
to $v_2$ and $\nu N_1$, $E^0 N_2$ masses to $v_4$.  Finally, 
$15_F \times 15_F$ transforms as $15_S^*$, with $u^c u$ masses proportional to 
$v_4$ as well.

Neutrinos obtain Majorana seesaw masses proportional to $v_4^2/v_2$, 
with $N_1$ acting as the usual right-handed neutrino in left-right models. 
This shows that $SU(6)$ may be used for seesaw neutrino masses in lieu of 
the customary $SO(10)$.  The $3 \times 3$ mass matrix spanning 
$(N_2,E^0,\bar{E}^0)$ is of the form 
\begin{equation}
{\cal M}_{NE} = \pmatrix{f_N v_2 & f_{NE} v_4 & 0 \cr f_{NE} v_4 & 0 & 
f_E v_1 \cr 0 & f_E v_1 & 0}.
\end{equation}

\noindent \underline{\it Gauge Boson Masses and Interactions}~:~
The gauge bosons belonging to the adjoint 35 representation of $SU(6)$ are 
superheavy with masses proportional to $v_3$ except for those corresponding 
to $SU(3)_C \times SU(2)_L \times U(1)_Y \times U(1)_N$.  The breaking of 
$U(1)_N$ comes from $v_{1,2}$ and that of 
$SU(2)_L \times U(1)_Y \times U(1)_N$ comes from $v_{4,5}$.  The charged 
$W^\pm$ mass is given by
\begin{equation}
m_W^2 = {1 \over 2} g_L^2 (v_4^2+v_5^2),
\end{equation}
the massless photon is $A=(e/g_L)W_3 + (e/g_Y)B$, where $B$ is the $U(1)_Y$ 
gauge boson and $g_Y/g_L=\tan \theta_W$ with $e^{-2} = g_L^{-2}+g_Y^{-2}$, 
whereas the $2 \times 2$ mass-squared matrix spanning $(Z,Z_N)$, where 
$Z = W_3 \cos \theta_W - B \sin \theta_W$, is given by
\begin{equation}
{\cal M}^2_{Z Z_N} = \pmatrix{(g_Z^2/2)(v_4^2+v_5^2) & g_Z g_N (4v_4^2-v_5^2) 
\cr g_Z g_N (4v_4^2-v_5^2) & 2g_N^2(25v_1^2+100v_2^2+16v_4^2+v_5^2)},
\end{equation} 
where $g_Z^2 = g_L^2 + g_Y^2$.  For simplicity, $v_5 = 2v_4$ may be assumed, 
so that $Z$ and $Z_N$ do not mix, thereby preserving all electroweak 
precision measurements involving the $Z$ boson.

The gauge interactions of $Z$ are given by
\begin{equation}
{\cal L}_Z = -g_Z Z_\mu j_Z^\mu = -g_Z Z_\mu (j_{3L}^\mu - \sin^2 \theta_W j_Q^\mu),
\end{equation}
and those of $Z_N$ by
\begin{eqnarray}
{\cal L}_N &=& -g_N {Z_N}_\mu j_{N}^\mu = -g_N {Z_N}_\mu [2\bar{u}_L \gamma^\mu 
u_L - 2\bar{u}_R \gamma^\mu u_R + 2 \bar{d}_L \gamma^\mu d_L + \bar{d}_R 
\gamma^\mu d_R \nonumber \\ && - \bar{e}_L \gamma^\mu e_L - 2 \bar{e}_R 
\gamma^\mu e_R - \bar{\nu}_L \gamma^\mu \nu_L + 5 \bar{N}_{1L} \gamma^\mu 
N_{1L} + 5 \bar{N}_{2L} \gamma^\mu N_{2L} - 4 \bar{D}_L \gamma^\mu D_L 
\nonumber \\ && + \bar{D}_R \gamma^\mu D_R - \bar{E}_L^- \gamma^\mu E_L^- 
+ 4 \bar{E}_R^- \gamma^\mu E_R^- - \bar{E}_L^0 \gamma^\mu E_L^0 
+ 4 \bar{E}_R^0 \gamma^\mu E_R^0].
\end{eqnarray}
As such, $Z_N$ may be produced at the collider through its couplings to $u$ 
and $d$ quarks, and be discovered through its couplings to charged leptons. 
The present collider limit~\cite{pdg18} is estimated to be a few TeV.

\noindent \underline{\it Electroweak Scalar Sector}~:~
At the level of $SU(3)_C \times SU(2)_L \times U(1)_Y \times U(1)_N$, there 
are two scalar singlets $\eta_1 \sim (1,1,0,-5)$, $\eta_2 \sim (1,1,0,10)$, 
and two scalar doublets $\Phi_4=(\phi_4^0,\phi_4^-) \sim (1,2,-1/2,4)$, 
$\Phi_5=(\phi_5^+,\phi_5^0) \sim (1,2,1/2,1)$.  Their Yukawa couplings with 
the fermions of (1) are
\begin{eqnarray}
&& D^c D \eta_1^*, ~~~ (E^- E^+ + E^0 \bar{E}^0) \eta_1^*, ~~~ 
N_1 N_1 \eta_2^*, ~~~ N_2 N_2 \eta_2^*, ~~~ 
d^c (d \bar{\phi}_5^0 + u \phi_5^-), \\ 
&& u^c (u \bar{\phi}_4^0 + d \phi_4^+), ~~~ 
e^c (e \bar{\phi}_5^0 + \nu \phi_5^-), ~~~  
N_1 (\nu \bar{\phi}_4^0 + e \phi_4^+), ~~~ N_2 (E^0 \bar{\phi}_4^0 + E^- 
\phi_4^+).
\end{eqnarray}
Note that these dimension-four terms all obey the imposed $Z_2$ symmetry 
discussed earlier under which $d^c$, $(\nu,e)$, and $(\phi_5^+,\phi_5^0)$ 
are odd, and the others are even.

The Higgs potential is
\begin{eqnarray}
V &=& m_1^2 \eta_1^* \eta_1 + m_2^2 \eta_2^* \eta_2 + m_4^2 \Phi_4^\dagger 
\Phi_4 + m_5^2 \Phi_5^\dagger \Phi_5 + [\mu_1 \eta_1 \tilde{\Phi}_4^\dagger 
\Phi_5 + \mu_2 \eta_1^2 \eta_2 + H.c.] \nonumber \\ 
&+& {1 \over 2} \lambda_1 (\eta_1^* \eta_1)^2 + {1 \over 2} \lambda_2 
(\eta_2^* \eta_2)^2  +{1 \over 2} \lambda_4 (\Phi_4^\dagger \Phi_4)^2 + 
{1 \over 2} \lambda_5 (\Phi_5^\dagger \Phi_5)^2 \nonumber \\ 
&+& \lambda_{12} (\eta_1^* 
\eta_1) (\eta_2^* \eta_2) + \lambda_{14} (\eta_1^* \eta_1) (\Phi_4^\dagger 
\Phi_4) + \lambda_{15} (\eta_1^* \eta_1) (\Phi_5^\dagger \Phi_5)  
+ \lambda_{24} (\eta_2^* \eta_2) (\Phi_4^\dagger \Phi_4) \nonumber \\ 
&+& \lambda_{25} 
(\eta_2^* \eta_2) (\Phi_5^\dagger \Phi_5) + \lambda_{45} (\Phi_4^\dagger 
\Phi_4) (\Phi_5^\dagger \Phi_5) + \lambda'_{45} (\Phi_4^\dagger \Phi_5) 
(\Phi_5^\dagger \Phi_4).
\end{eqnarray}
Note that the dimension-three $\mu_1$ term breaks the $Z_2$ symmetry softly. 
Together with the $\mu_2$ term, they ensure that there would be no extra 
accidental U(1) symmetry in $V$ beyond $U(1)_Y$ and $U(1)_N$.

The minimum of $V$ is determined by
\begin{eqnarray}
0 &=& v_1(m_1^2 + \lambda_1 v_1^2 + \lambda_{12} v_2^2 + \lambda_{14} v_4^2 
+ \lambda_{15} v_5^2 + 2 \mu_2 v_2) - \mu_1 v_4 v_5, \\ 
0 &=& v_2(m_2^2 + \lambda_2 v_2^2 + \lambda_{12} v_1^2 + \lambda_{24} v_4^2 
+ \lambda_{25} v_5^2) + \mu_2 v_1^2, \\ 
0 &=& v_4(m_4^2 + \lambda_4 v_4^2 + \lambda_{14} v_1^2 + \lambda_{24} v_2^2 
+ \lambda_{45} v_5^2) - \mu_1 v_1 v_5, \\ 
0 &=& v_5(m_5^2 + \lambda_5 v_5^2 + \lambda_{15} v_1^2 + \lambda_{25} v_2^2 
+ \lambda_{45} v_4^2) - \mu_1 v_1 v_4. 
\end{eqnarray}
The $4 \times 4$ mass-squared matrix spanning $\sqrt{2}Im(\eta_1^0,\eta_2^0,
\phi_4^0,\phi_5^0)$ is given by
\begin{equation}
{\cal M}_A^2 = \pmatrix{\mu_1 v_4 v_5/v_1 - 4 \mu_2 v_2 & -2 \mu_2 v_1 & 
\mu_1 v_5 & \mu_1 v_4 \cr -2 \mu_2 v_1 & - \mu_2 v_1^2/v_2 & 0 & 0 \cr 
\mu_1 v_5 & 0 & \mu_1 v_1 v_5/v_4 & \mu_1 v_1 \cr \mu_1 v_4 & 0 & \mu_1 v_1 
& \mu_1 v_1 v_4/v_5}.
\end{equation}
Two zero eigenvalues appear, corresponding to 
$[v_1,-2v_2,-v_4 v_5^2/(v_4^2+v_5^2),-v_4^2 v_5/(v_4^2+v_5^2)]$ 
and $(0,0,v_4,-v_5)$, becoming 
the longitudinal components of $Z_N$ and $Z$ respectively. The remaining 
two massive pseudoscalar components span $(2v_2,v_1,0,0)$ and 
$[v_1,-2v_2,(v_1^2+4v_2^2)/v_4,(v_1^2+4v_2^2)/v_5]$ with $2 \times 2$ 
mass-squared matrix
\begin{equation}
\pmatrix{4 \mu_1 v_2^3 v_4^2 v_5^2 - \mu_2 v_1 
(v_1^2+4v_2^2)^2 v_4 v_5 &  2 \mu_1 v_1 v_2^2 v_4 v_5 
\sqrt{v_4^2v_5^2 + (v_1^2+4v_2^2)(v_4^2+v_5^2)} \cr 
2 \mu_1 v_1 v_2^2 v_4 v_5 \sqrt{v_4^2v_5^2 + (v_1^2+4v_2^2)
(v_4^2+v_5^2)}  & \mu_1 v_1^2 v_2 [v_4^2v_5^2 + (v_1^2+4v_2^2)(v_4^2+v_5^2)]},
\end{equation}
divided by $v_1 v_2 (v_1^2+4v_2^2) v_4 v_5$.  It shows explicitly that 
$\mu_1=0$ or $\mu_2=0$ implies one zero eigenvalue, and $\mu_1=\mu_2=0$ 
implies two.  In the limit $v_{4,5} << v_{1,2}$, it reduces to
\begin{equation}
{\cal M}^2_A = \pmatrix{-\mu_2(v_1^2+4v_2^2)/v_2 & 0 \cr 0 & \mu_1 v_1 
(v_4^2 + v_5^2)/v_4 v_5}.
\end{equation}

The $4 \times 4$ mass-squared matrix spanning $\sqrt{2}Re(\eta_1^0,\eta_2^0,
\phi_4^0,\phi_5^0)$ is given by
\begin{equation}
\pmatrix{2 \lambda_1 v_1^2 + \mu_1 v_4 v_5/v_1 -4 \mu_2 v_2& 
2 \lambda_{12} v_1 v_2 + 2 \mu_2 v_1 & 2 \lambda_{14} v_1 v_4 - \mu_1 v_5 & 
2 \lambda_{15} v_1 v_5 - \mu_1 v_4 \cr 2 \lambda_{12} v_1 v_2 + 2 \mu_2 v_1 & 
2 \lambda_2 v_2^2 - \mu_2 v_1^2/v_2 & 2 \lambda_{24} v_2 v_4 & 2 \lambda_{25} 
v_2 v_5 \cr 2 \lambda_{14} v_1 v_4 - \mu_1 v_5 & 2 \lambda_{24} v_2 v_4 & 
2 \lambda_4 v_4^2 + \mu_1 v_1 v_5/v_4 & 2 \lambda_{45} v_4 v_5 - \mu_1 v_1 
\cr 2 \lambda_{15} v_1 v_5 - \mu_1 v_4 & 2 \lambda_{25} v_2 v_5 & 2 
\lambda_{45} v_4 v_5 - \mu_1 v_1 & 2 \lambda_5 v_5^2 + \mu_1 v_1 v_4/v_5}.
\end{equation}
Let $h = \sqrt{2}[v_4 Re(\phi_4^0) + v_5 Re(\phi_5^0)]/\sqrt{v_4^2+v_5^2}$, 
then its mass is given by
\begin{equation}
m_h^2 = {2\lambda_4 v_4^4 + 2\lambda_5 v_5^4 + 4 \lambda_{45} v_4^2 v_5^2 
\over v_4^2 + v_5^2}.
\end{equation}
It is the only linear combination of the four neutral scalar fields which 
has no $v_{1,2}$ contribution to its mass, and acts as the SM Higgs boson 
in its interactions.   The other three scalar bosons are much heavier and 
have suppressed mixing with $h$, assuming again $v_{4,5} << v_{1,2}$,

Consider now the linear combinations 
$S_1 = \sqrt{2}[v_1 Re(\eta_1^0) + 2 v_2 Re(\eta_2^0)]/\sqrt{v_1^2+4v_2^2}$ and 
$S_2 = \sqrt{2}[2 v_2 Re(\eta_1^0) - v_1 Re(\eta_2^0)]/\sqrt{v_1^2+4v_2^2}$. 
Then
\begin{eqnarray}
m^2_{S_1} &=& {2 \lambda_1 v_1^4 + 8 \lambda_2 v_2^4 + 8 \lambda_{12} v_1^2 v_2^2 
+ \mu_1 v_1 v_4 v_5 \over v_1^2 + 4 v_2^2}, \\ 
m^2_{S_2} &=& - {\mu_2 \over v_2} (v_1^2+4v_2^2) + (8 \lambda_1 + 2 \lambda_2 - 
8 \lambda_{12}) {v_1^2 v_2^2 \over v_1^2 + 4v_2^2} + {4\mu_1 v_4 v_5 v_2^2 
\over v_1 (v_1^2 + 4 v_2^2)}, \\ 
m^2_{S_1 S_2} &=& {2 v_1 v_2 \over v_1^2 + 4 v_2^2} [(2 \lambda_1 - 
\lambda_{12})v_1^2 + (4\lambda_{12}-2\lambda_2)v_2^2] + {2 \mu_1 v_4 v_5 v_2 \over 
v_1^2 + 4v_2^2}.
\end{eqnarray}
This shows that if $\mu_2 >> v_{1,2}$, then $S_2$ is much heavier than 
$S_1$ and their mixing is supressed. This scenario is useful for the 
dark matter phenomenology to be discussed later.

The remaining scalar 
$H = \sqrt{2}[v_5 Re(\phi_4^0) - v_4 Re(\phi_5^0)]/\sqrt{v_4^2+v_5^2}$ 
has mass given by
\begin{equation}
m^2_{H} = {\mu_1 v_1 (v_4^2+v_5^2) \over v_4 v_5} + {(2 \lambda_4 + 2 \lambda_5 - 
4 \lambda_{45}) v_4^2 v_5^2 \over v_4^2 + v_5^2}.
\end{equation}
It mixes with $h$ through the term
\begin{equation}
m^2_{hH} = {2v_4 v_5 [(\lambda_4-\lambda_{45})v_4^2 + 
(\lambda_{45}-\lambda_5)v_5^2] \over v_4^2+v_5^2}.
\end{equation}
This shows that $\mu_1, v_1 >> v_{4,5}$ guarantees that $H$ is heavier than $h$ 
and their mixing is suppressed, as remarked earlier.

\noindent \underline{\it Dark Matter}~:~
Of the new particles beyond those of the Standard Model (SM), $N_1$ acts 
as the seesaw anchor of $\nu$ as already explained.  It replaces the usually 
assumed right-handed neutrino.  The others are the color-triplet fermions 
$D$ of charge $-1/3$, the vectorlike electroweak doublet fermions 
$(E^-,E^0)$ and the neutral singlet $N_2$ fermion.  At the level of the 
SM extension to $U(1)_N$, as is clear from (7) and (8), this latter 
set of particles are distinguished from those of the SM by a discrete 
$Z_2$ symmetry under which they are odd.  The lightest, presumably $N_2$, 
could then be dark matter.  This is analogous to the stability of the 
proton from baryon number conservation in the SM.  However, once it is 
realized that these particles are embedded into $SU(6)$, it is clear 
that $N_2$ must also decay, just as the proton.

The superheavy gauge bosons $(5,6)$ and $(5^*,-6)$ of the adjoint vector 35 
representation connect $N_1$ to $(d^c,e,\nu)$, $N_2$ to $(D^c,E^-,E^0)$, 
and $(D,E^+,\bar{E}^0)$ to $(u^c,u,d,e^c)$, as shown in (1). 
Together with (2), they allow the decay $N_2 \to e^+ W^-$ through 
the superheavy neutral gauge boson $X^0$ in $(5,6)$ as shown in Fig.~1.
\begin{figure}[htb]
\vspace*{-5cm}
\hspace*{-3cm}
\includegraphics[scale=1.0]{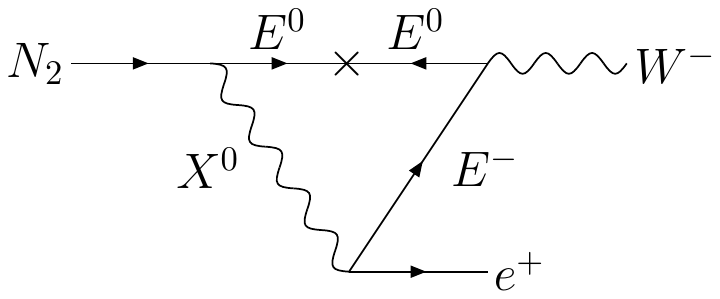}
\vspace*{-21.5cm}
\caption{Decay of $N_2$ through superheavy gauge boson $X^0$.}
\end{figure}
This amplitude is proportional to $(v_4^2/v_2)(m_E/m_X^2)$, and is of 
similar magnitude to that of proton decay.  This establishes the notion 
that dark matter stability is akin to proton stability in the context 
of the unification of matter and dark matter, in parallel to that of quarks 
and leptons.

As for the heavy $E$ leptons, it is clear that $E^0$ decays to 
$N_2 h$ from (8) and $E^-$ decays to $E^0 W^-$.  The heavy $D$ quark 
decays through the scalar $(3,1,-1/3,-4)$ component of $(5,-4)$ in $15_S$ 
or $21_S$ to $N_2 d N_1$, with $N_1$ decaying to $\nu h$.  Note that these 
interactions do not violate the $Z_2$ symmetry separating matter from 
dark matter at low energy.  They serve the purpose of allowing the heavier 
dark particles to decay to $N_2$ rapidly, assuming that the mediating scalars 
are not too heavy.  Note also that proton decay is possible through scalar 
exchange as in $SU(5)$.  Here it occurs through the mixing of $(3,1,-1/3,-4)$ 
with $(3,1,-1/3,1)$ through the term $6_S \times 15^*_S \times 84_S$, and 
may be suppressed with a large mass for $(3,1,-1/3,1)$ as in $SU(5)$.

\noindent \underline{\it Relic Abundance and Direct Search}~:~
The relic abundance of the very long-lived $N_2$ is determined by its 
annihilation to scalar bosons which are in thermal equilibrium with SM 
particles.  In particular, the dominant process is shown in Fig.~2, 
assuming $m_{S_1} < m_{N_2} << m_{S_2}$.
\begin{figure}[htb]
\vspace*{-5cm}
\hspace*{-3cm}
\includegraphics[scale=1.0]{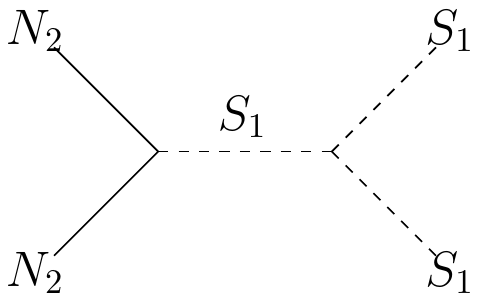}
\vspace*{-21.5cm}
\caption{$N_2 N_2$ annihilation to $S_1 S_1$.}
\end{figure}
The $N_2 N_2 S_1$ coupling is $\sqrt{2}m_{N_2}/\sqrt{v_1^2+4v_2^2}$, and 
the $S_1 S_1 S_1$ coupling is dominated by 
$6 \sqrt{2} \mu_2 v_1^2 v_2/(v_1^2+4v_2^2)^{3/2}$.  Hence the annihilation 
cross section at rest multiplied by relative velocity is 
\begin{equation}
\sigma_{ann} \times v_{rel} = {9 \mu_2^2 v_1^4 v_2^2 \over 128 \pi m_{N_2}^2 
(v_1^2 + 4 v_2^2)^4} \sqrt{ 1-{m^2_{S_1} \over m_{N_2}^2}} \left( 1 - 
{m^2_{S_1} \over 4m_{N_2}^2} \right)^{-2}.
\end{equation}
As an example, let $v_1=2v_2=5$ TeV, $m_{N_2} = 1$ TeV, $m_{S_1} = 800$ GeV, 
and $\mu_2 = 14.7$ TeV, then 
$\sigma_{ann} \times v_{rel} = 3 \times 10^{-26}~{\rm cm}^3/{\rm s}$, the 
canonical value for the correct dark matter relic abundance of the Universe.

As for direct search, since $N_2$ is a Majorana fermion, it does not 
couple to the $Z_N$ gauge boson at rest, so its only interaction with 
matter at underground experiments is through the SM Higgs boson. 
The mixing of $S_1$ with $h$ comes from the term
\begin{equation}
m^2_{S_1 h} = {2v_1^2(\lambda_{14}v_4^2+\lambda_{15}v_5^2) + 
2v_2^2(\lambda_{24}v_4^2+\lambda_{25}v_5^2) - 2 \mu_1 v_1 v_4 v_5 
\over \sqrt{v_1^2+4v_2^2} \sqrt{v_4^2+v_5^2}}.
\end{equation}
Let $\theta = m^2_{S_1 h}/m^2_{S_1}$, then the coupling of $N_2$ to $h$ is 
$m_N \theta \sqrt{2/(v_1^2+4v_2^2)}$, whereas $h$ couples to quarks by 
$m_q/\sqrt{2(v_4^2+v_5^2)}$.  The spin-independent elastic scattering 
cross section of $N _2$ off a Xenon nucleus per nucleon is given by
\begin{equation}
\sigma_0 = {4 \over \pi} \left( {m_{N_2} m_{Xe} \over m_{N_2} + m_{Xe}} 
\right)^2 \left| {54 f_p + 77 f_n \over 131} \right|^2,
\end{equation}
where~\cite{hint11}
\begin{eqnarray}
{f_p \over m_p} &=& \left[0.075 + {2 \over 27}(1-0.075) \right] 
{m_{N_2} \theta \over m_h^2 \sqrt{v_1^2+4v_2^2}\sqrt{v_4^2+v_5^2}}, \\
{f_n \over m_n} &=& \left[0.078 + {2 \over 27}(1-0.078) \right]
{m_{N_2} \theta \over m_h^2 \sqrt{v_1^2+4v_2^2}\sqrt{v_4^2+v_5^2}}.
\end{eqnarray}
For $m_{N_2} = 1$ TeV, $v_1=2v_2=5$ TeV, $\sqrt{v_4^2+v_5^2}=174$ GeV, 
$m_h = 125$ GeV, the upper limit on $\theta$ from~\cite{xenon18} 
$\sigma_0 < 10^{-45}$ cm$^2$ is $1.84 \times 10^{-3}$.  Without fine tuning, 
$\theta$ is of order $\sqrt{(v_4^2+v_5^2)/(v_1^2+4v_2^2)} \sim 1/60$.  
Hence a fine tuning of order 1/10 is required.  This is possible by 
adjusting the value of $\mu_1$ in the range of $v_{1,2}$ relative to 
other free parameters in (25).

\noindent \underline{\it Gauge Coupling Unification}~:~
The fermion content of this model is that of the SM extended by two 
complete $SU(5)$ fermion multiplets $(D^c,D^c,D^c,E^-,E^0)$, 
$(D,D,D,E^+,\bar{E}^0)$, and two fermion singlets $N_{1,2}$ per family.  
There are also two electroweak Higgs doublets $\Phi_{4,5}$ instead of 
just one in the SM, and two Higgs singlets.  As far as the evolution of 
the known three gauge couplings, it follows the same pattern as the SM 
model with two Higgs doublets.  It is well-known that these gauge couplings 
do not unify under such circumstances.  However, with the appropriate 
addition of some particle multiplets~\cite{bm84,k93,gl03,m05}, unification 
of gauge couplings is possible at a high energy scale. 

Consider the one-loop renormalization-group equations
\begin{equation}
{1 \over \alpha_i(M_1)} - {1 \over \alpha_i(M_2)} = {b_i \over 2 \pi} \ln 
{M_2 \over M_1},
\end{equation}
where $\alpha_i = g_i^2/4\pi$ and the coefficients $b_i$ are determined by 
the particle content between $M_1$ and $M_2$.  In the SM with one Higgs 
doublet, these are given by
\begin{eqnarray}
SU(3)_C &:& b_C = -11 + (4/3)N_F = -7, \\ 
SU(2)_L &:& b_L = -22/3 + (4/3)N_F + 1/6 = -19/6, \\ 
U(1)_Y &:& b_Y = (4/3)N_F + 1/10,
\end{eqnarray}
where $N_F=3$ is the number of quark and lepton families and $b_Y$ has been 
normalized by the well-known factor of 3/5.

The contributions of the extra fermions and scalars of this model supply 
the following changes
\begin{equation}
\Delta b_C = (2/3)N_F, ~~~ \Delta b_L = (2/3)N_F + 1/6, ~~~ \Delta b_Y = 
(2/3)N_F + 1/10.
\end{equation}
The further addition of the $\xi \sim (3,1,-1/3,-4)$ scalar contributes 
$\Delta b_C = 1/6$ and $\Delta b_Y = 1/15$.

Gauge coupling unification may be achieved following Ref.~\cite{m18}, 
by adding a colored fermion octet $\Omega \sim (8,1,0,0)$ with 
$\Delta b_C = 2$ as well as an electroweak fermion triplet 
$\Sigma \sim (1,3,0,0)$ with $\Delta b_L = 4/3$, both from 
an assumed $35_F$ of $SU(6)$, and a scalar triplet $S \sim (1,3,0,-5)$ 
with $\Delta b_L = 2/3$ from the $84_S$ already present.  The resulting 
evolution equations become
\begin{eqnarray}
{1 \over \alpha_U} &=& {1 \over \alpha_C} + {17/6 \over 2\pi} \ln {M_U 
\over M_Z} + {1/6 \over 2\pi} \ln {M_\xi \over M_Z} + {2 \over 2\pi} \ln 
{M_\Omega \over M_Z} + {2 \over 2\pi} \ln {M_{Z_N} \over M_Z}, \\ 
{1 \over \alpha_U} &=& {1 \over \alpha_L} - {1 \over 2\pi} \ln {M_U 
\over M_Z} + {4/3 \over 2\pi} \ln {M_\Sigma \over M_Z} + {2/3 \over 2\pi} \ln 
{M_S \over M_Z} + {2 \over 2\pi} \ln {M_{Z_N} \over M_Z} + {1/6 \over 2\pi} 
\ln {M_\Phi \over M_Z}, \\
{1 \over \alpha_U} &=& {3 \over 5\alpha_Y} - {94/15 \over 2\pi} \ln {M_U 
\over M_Z} + {1/15 \over 2\pi} \ln {M_\xi \over M_Z} + {2 \over 2\pi} 
\ln {M_{Z_N} \over M_Z} + {1/10 \over 2\pi} \ln {M_\Phi \over M_Z},
\end{eqnarray}
where $M_\Phi$ is the mass of the heavier scalar doublet, and 
$\alpha_C, \alpha_L, \alpha_Y$ are evaluated at $M_Z$, with central 
values given by~\cite{pdg18}
\begin{equation}
\alpha_C = 0.118, ~~~ \alpha_L = (\sqrt{2}/\pi)G_F M_W^2 = 0.0340, ~~~ 
\alpha_Y = \alpha_L \tan^2 \theta_W = 0.0102.
\end{equation}
Eliminating $\alpha_U$, the two conditions on the various intermediate 
masses are
\begin{eqnarray}
34.318 &=& \ln {M_U \over M_Z} + {1 \over 23} \ln {M_\xi \over M_Z} + 
{12 \over 23} \ln {M_\Omega \over M_Z} - {8 \over 23} \ln {M_\Sigma \over M_Z} 
- {4 \over 23} \ln {M_S \over M_Z} - {1 \over 23} \ln {M_\Phi \over M_Z}, \\ 
35.089 &=& \ln {M_U \over M_Z} - {1 \over 79} \ln {M_\xi \over M_Z} + 
{20 \over 79} \ln {M_\Sigma \over M_Z} + {10 \over 79} \ln {M_S \over M_Z} 
+ {1 \over 79} \ln {M_\Phi \over M_Z}.
\end{eqnarray}
Subtracting the two equations to eliminate $M_U$, and assuming 
$M_\Omega = M_\Sigma = M_S$, the condition
\begin{equation}
0.771 = -{102 \over (79)(23)} \ln {M_\xi \over M_Z} + {30 \over 79} \ln 
{M_S \over M_Z} + {102 \over (79)(23)} \ln {M_\Phi \over M_Z}
\end{equation}
is obtained.  This is satisfied for example with $M_\Phi = 500$ GeV, 
$M_S = 1$ TeV, and $M_\xi = 5.9$ TeV, resulting in $M_U = 6.57 \times 10^{16}$ 
GeV and $\alpha_U = 0.0386$ for $M_{Z_N} = 3$ TeV.  As for $\alpha_N$, 
using the normalization factor of $1/60$, it may now be deduced from
\begin{equation}
{1 \over \alpha_U} = {1 \over 60 \alpha_N} - {227/30 \over 2\pi} \ln 
{M_U \over M_Z} + {4/15 \over 2\pi} \ln {M_\xi \over M_Z} + {5/12 \over 2\pi} 
\ln {M_S \over M_Z} + {187/36 \over 2\pi} \ln {M_{Z_N} \over M_Z} + 
{17/180 \over 2\pi} \ln {M_\Phi \over M_Z},
\end{equation}
yielding $\alpha_N = 2.61 \times 10^{-4}$.

\noindent \underline{\it Conclusion}~:~
Following up on the notion~\cite{m13} that dark matter is a long-lived 
particle just as the proton in a grand unified theory by extending $SU(5)$ 
to $SU(6)$, a specific complete model is presented.  The SM fermions are 
augmented per family by one heavy $D$ quark, one heavy vectorlike lepton 
doublet $(E^0,E^-)$, and two singlets $N_{1,2}$, as shown in (1).  A discrete 
$Z_2$ symmetry is imposed on the dimension-four terms of the Lagrangian, but 
is softly and spontaneously broken in the scalar sector. The resulting theory 
has the following features.  The heavy Majorana fermion $N_1$ acts as the 
right-handed neutrino in allowing the corresponding observed neutrino to have a 
small seesaw mass, doing this in the context of $SU(6)$ instead of $SO(10)$. 
The heavy Majorana fermion $N_2$ is a dark-matter candidate with interactions 
consistent with the correct relic abundance and may be observed in underground 
direct-search experiments through its induced coupling to the SM Higgs boson. 
However, $N_2$ is not protected by a symmetry and decays through a superheavy 
gauge boson in $SU(6)$, just as the proton is not absolutely stable and 
decays through a superheavy gauge boson in $SU(5)$.  The pattern of symmetry 
breaking is assumed to be 
$SU(6) \to SU(3)_C \times SU(2)_L \times U(1)_Y \times U(1)_N$ at a scale 
of order $10^{16}$ GeV.  Gauge coupling unification is possible and 
demonstrated with an explicit example.  The heavier $D$ and $(E^0,E^-)$ 
particles decay rapidly to $N_2$ through scalars at an intermediate mass. 
The $Z_N$ gauge boson is potentially observable at a few TeV.

By incorporating dark matter as an essential component of grand unification, 
it is shown that just as the conservation of baryon number and lepton parity 
in the SM is violated in the unification of quarks and leptons, the 
conservation of dark parity at low energy is violated in the unification of 
matter and dark matter.  The longevity of dark matter is then linked to the 
longevity of the proton, as a natural explanation of the former's existence.

\noindent \underline{\it Acknowledgement}~:~
This work was supported in part by the U.~S.~Department of Energy Grant 
No. DE-SC0008541.

\bibliographystyle{unsrt}

\end{document}